\documentclass[11pt,a4paper]{article}

\usepackage[dvips,bookmarks,colorlinks=true,linkcolor=blue,urlcolor=red]{hyperref}

\usepackage{amsmath,amsfonts,amsthm}

\newcommand{\cH}{\mathcal{H}}

\newcommand{\RR}{\mathbb{R}}
\newcommand{\ZZ}{\mathbb{Z}}

\newcommand{\EE}{\mathbb{E}}
\newcommand{\PP}{\mathbb{P}}
\newcommand{\NN}{\mathbb{N}}

\newcommand{\cE}{\mathcal{E}}
\newcommand{\cV}{\mathcal{V}}

\newcommand{\tr}{\mathop{\mathrm{tr}}}

\newcommand{\spec}{\mathop{\mathrm{spec}}}

\newcommand{\dist}{\mathop{\mathrm{dist}}}

\newcommand{\mi}{\mathrm{min}}
\newcommand{\ma}{\mathrm{max}}

\newtheorem{theorem}{Theorem}

\theoremstyle{definition}

\newtheorem{rem}[theorem]{\bf Remark}

\begin{document}

\title{\bf Localization on quantum graphs\\
  with random edge lengths}

\author{\LARGE Fr{\'e}d{\'e}ric Klopp ${}^{1,2}$ \and \LARGE Konstantin Pankrashkin ${}^{1,3,4}$ \\[\bigskipamount]
  ${}^1$ Laboratoire Analyse, G{\'e}om{\'e}trie et Applications, CNRS UMR 7539\\
  Institut Galil{\'e}e, Universit{\'e} Paris-Nord\\ 99 av. Jean-Baptiste Cl{\'e}ment, 93430 Villetaneuse, France\\[\smallskipamount]
  ${}^2$ Institut Universitaire de France\\
  103 bd. Saint-Michel, 75005 Paris, France\\
  E-mail: \texttt{klopp@math.univ-paris13.fr}\\[\smallskipamount]
  ${}^3$ On leave from: Institut f{\"u}r Mathematik, Humboldt-Universit{\"a}t
  \\ Rudower
  Chaussee 25, 12489 Berlin, Germany\\[\smallskipamount]
  ${}^4$ Current address: Laboratoire de math{\'e}matiques, CNRS UMR 8628\\
  Unversit{\'e} Paris-Sud,  B{\^a}timent 425, 91405 Orsay, France\\
  E-mail: \texttt{konstantin.pankrashkin@math.u-psud.fr} }

\date{}

\maketitle

\begin{abstract}
  \noindent The spectral properties of the Laplacian on a class of
  quantum graphs with random metric structure are studied.  Namely, we
  consider quantum graphs spanned by the simple $\ZZ^d$-lattice with
  $\delta$-type boundary conditions at the vertices, and we assume
  that the edge lengths are randomly independently identically
  distributed. Under the assumption that the coupling constant at the
  vertices does not vanish, we show that the operator exhibits the
  Anderson localization near the spectral edges situated outside
  a certain forbidden set.
\end{abstract}


\bigskip




\section{Introduction}
\label{sec:introduction}

In the present paper we are discussing the Anderson localization for a
special class of random perturbations of periodic structures.  As
argued in the original paper by Anderson \cite{An}, there is no
transport in disordered media under certain conditions.  This
phenomenon was interpreted initially within the framework of spectral
theory: as proved in various settings, a generic random perturbation
of periodic operators produces a dense pure point spectrum in certain
energy intervals; at the same time one can be interested in the
dynamical localization consisting in uniform bounds, in both space and
energy, for propagating wave packets, which implies the dense pure
point spectrum.  We refer to the very recent collection of papers
\cite{EDR} providing the state of art in the theory of random
Schr{\"o}dinger operators, in particular, the paper \cite{Klein}
discussing various mathematical interpretations of the Anderson
localization.

Up to now, the most studied models of disordered media are either
discrete (tight-binding) Hamiltonians or continuous Schr{\"o}dinger
operators.  In the last decade, there is an increasing interest in the
analysis of quantum Hamiltonians on so-called quantum graphs, i.e.
differential operators on singular one-dimensional spaces, see the
collection of papers~\cite{QG1,EDAG,AGA,Ku}. A quantum graph is
composed of one-dimensional differential operators on the edges and
boundary conditions at the vertices describing coupling of edges.
Such operators provide an effective model for the study of various
phenomena in the condensed matter physics admitting an experimental
verification, and there is natural question about the influence of
random perturbations in such systems~\cite{Vid}. There are numerous
possibilities to introduce randomness: combinatorial structure,
coefficients of differential expression, coupling, metric, etc. Being
locally of one-dimensional nature and admitting a complex global
shape, quantum graphs take an intermediate position between the
one-dimensional and higher dimensional Schr{\"o}dinger operators.  It
seems that the paper \cite{KSch} considering the random necklace model
was the first one discussing random interactions and the Anderson
localization in the quantum graph setting. Later, these results were
generalized for radial tree configurations \cite{HP}, where Anderson
localization at all energies was proved. Both papers used a machinery
specific for one-dimensional operators. The paper \cite{ASW} addressed
the spectral analysis on quantum tree graphs with random edge lengths;
it appears that the Anderson localization does not hold near the
bottom of the spectrum at least in the weak disorder limit and one
always has some absolutely continuous spectrum.  Another important
class of quantum graphs is given by $\ZZ^d$-lattices.  The paper
\cite{EHS} studied the situation where each edge carries a random
potential and showed the Anderson localization near the bottom of the
spectrum. Some generalizations were then obtained in \cite{GLV,GHV}.
The case of random coupling was considered by the present authors in
\cite{FKP}; recently we learned on an earlier paper \cite{CM} where
some preliminary estimates for the same model were obtained.

The present Letter is devoted to the study of quantum graphs spanned
by the $\ZZ^d$-lattice where the edge lengths are random independent
identically distributed variables. We consider the free Laplacian on
each edge with $\delta$-type boundary conditions and show, under
certain technical assumptions, that the operator exhibits the Anderson
localization at the bottom of the spectrum, i.e. that the bottom of
the spectrum is pure point with exponentially decaying eigenfunctions.

There are two basic methods of proving localization for random
operators: the multiscale analysis going back to Fr{\"o}hlich and Spencer
\cite{FS} and the Aizenman-Molchanov method \cite{AM}. The
Aizenman-Molchanov method gives explicit and efficient criteria for
localization in terms of the Green function but only works under
special assumptions on the way the randomness enters the problem (we
used this method for the study of the random coupling model in
\cite{FKP}), which does not hold in the situation we are studying. On
the other hand, the multiscale analysis is a rather universal tool
which can handle very abstract situations \cite{St}.  It is a certain
iterative procedure which works as far as some input data are
available (see the paper \cite{EHS} discussing the multiscale analysis
for quantum graphs).  Below we are concentrating on obtaining the
the most important necessary ingredients, more precisely, the Wegner estimate and the
initial scale estimate.  We are mostly interested in
the spectral interpretation of the Anderson localization, and our
results, being combined with the multiscale analysis, prove the presence of
the dense pure point spectrum in respective energy ranges.
Nevertheless, they also can be used for the study of the dynamical
localization; we refer to \cite{Klein} for details.

During the revision of the Letter the preprint
\cite{LPPV} appeared, which studies a similar model (but with a different problem setting)
and contains an alternative proof of the Wegner estimate for the zero coupling constants.

\section{Random length model on a quantum graph lattice}\label{ss1}
\label{sec:random-length-model}

We recall here some basic constructions for quantum graphs.  For the
general theory see e.g. the reviews \cite{GS,Ku1,Ku2} and the
collections of papers \cite{QG1,EDAG,AGA,Ku}.

Let $\Gamma=(\cV,\cE)$ be a countable directed graph with $\cV$ and
$\cE$ being the sets of vertices and edges, respectively.
For an edge $e\in \cE$, we denote by $\iota{e}$ its initial vertex and
by $\tau{e}$ its terminal vertex.  For $e\in\cE$ and $v\in \cV$ we
write $v\sim e$ or $e\sim v$ if $v\in\{\iota e,\tau e\}$. The degree
of a vertex is the number $\deg v:=\#\{e\in\cE:\,e\sim v\}$.

For $0<l_\mi<l_\ma<\infty$ consider a function $l:\cE\to
[l_\mi,l_\ma]$.  Sometimes we will write $l_e$ instead of $l(e)$; this
number will be interpreted as the length of the edge $e$.  Replacing
each edge $e$ by a copy of the segment $\big[0,l_e\big]$ in such a way
that $0$ corresponds to $\iota e$ and $l_e$ to $\tau e$, one
obtains a so-called metric graph. Our aim is to study a special-type
differential operator on such a structure.

In the space $\cH:=\bigoplus_{e\in\cE} \cH_e$, $\cH_e=L^2[0,l_e]$
consider an operator $H$ acting as $f=:(f_e)\mapsto(-f''_e)=:Hf$ on
the domain consisting of the functions $f\in \bigoplus_{e\in\cE}
H^2[0,l_e]$ satisfying the Kirchhoff boundary conditions, i.e. for any
$v\in\cV$ one has
\begin{equation}
  \label{eq-cont1}
  f_e(l_e)=f_b(0)=:f(v), \quad \tau e=\iota b=v,
\end{equation}
and
\begin{equation}
  \label{eq-cont2}
  f'(v)=\alpha f(v), \quad
  f'(v):=\sum_{e:\iota e=v} f'_e(0) - \sum_{e:\tau e=v} f'_e(l_e),
\end{equation}
where $\alpha$ is a real number, the so-called coupling constant (for
simplicity, we assume that the coupling constants are the same for all
vertices, which is sufficient for our purposes).  It is known that the
operator thus obtained is self-adjoint \cite{Ku1}.  We denote this
operator by $H(\Gamma,l,\alpha)$.

We are going to study a special case of underlying combinatorial configuration, namely a
periodic lattice with random edge lengths. Let $\cV=\ZZ^d$, $d\ge 1$, and $h_j$, $j=1,\dots,d$,
be the canonical basis of $\ZZ^d$. Set
\begin{equation*}
  \cE_d:=\{(m,m+h_j), \quad m\in\ZZ^d, \quad j=1,\dots,d\},
\end{equation*}
where for an edge $e=(v,v')\in \ZZ^d$, $v,v'\in\ZZ^d$, one has $\iota
e:=v$, $\tau e:=v'$.  For this graph $\Gamma^d:=(\ZZ^d,\cE_d)$
consider the operator $H(l,\alpha):=H(\Gamma^d,l,\alpha)$.  In the
present paper, we study some spectral properties of such operators
under the assumption that $l_e$ are random independent identically
distributed variables.

Namely, on $(\Omega,\PP)$, a probability space, let
$\big(l_e^\omega\big)_{e\in\cE}$ be a family of independent
identically distributed (i.i.d.) random variables whose common
distribution has a Lipschitz continuous density $\rho$ with support
$[l_\mi,l_\ma]$.  By a random Hamiltonian on the quantum graph, we
mean the family of operators $H^\omega(\alpha):=H(l^\omega,\alpha)$.

For $n\in\ZZ^d$ consider the shifts $\tau_n$ acting on the set of
edges, $\tau_n(m,m')=(m+n,m'+n)$ and the operators
$H(l^\omega_n,\alpha)$ with $l^\omega_n(e):=l^\omega(\tau_n e)$.
Clearly, $H(l^\omega_n,\alpha)$ is unitarily equivalent to
$H(l^\omega,\alpha)$ for any $n$ as $H(l^\omega_n,\alpha)U_n
=U_nH(l^\omega_n,\alpha)$, $U_n (f_e)=(f_{\tau_n e})$.  In terms of
the theory of random operators, the shifts $\tau_n$ form a measure
preserving ergodic family on $\Omega$ which allows one to use the
standard results from the theory of random operators~\cite{CL,PF}.  In
particular, one obtains the non-randomness of the spectrum and the
spectral components: there exist closed subsets
$\Sigma_j=\Sigma_j(\alpha)\subset\RR$ and a subset
$\Omega'\subset\Omega$ with $\PP(\Omega')=1$ such that $\spec_j
H^{\omega}(\alpha)=\Sigma_j$, $j\in\{\text{pp},\text{ac},\text{sc}\}$,
for any $\omega\in\Omega'$. We recall that the pure point spectrum
$\spec_\text{pp} H$ is the closure of the set of the eigenvalues of
$H$).  Let
$\Sigma(\alpha)=\Sigma_\text{pp}\cup\Sigma_\text{ac}\cup\Sigma_\text{sc}$
be the almost sure spectrum of $H^\omega(\alpha)$.  The aim of the
paper is to show that there are intervals $I\subset \Sigma$ such that
$\Sigma_\text{ac}\cap I=\Sigma_\text{sc}\cap I=\emptyset$; this means
that the interval $I$ is densely filled with
eigenvalues of $H^\omega$ and there is no continuous spectrum in $I$ for almost all
$\omega$, or, in other words, the spectrum of $H^\omega$ in $I$ is \emph{almost surely
dense pure point.}

Let us describe first the almost sure spectrum as a set.  For $u>0$
and $\beta\in\RR$ denote by $P_{l,\beta}$ the periodic Kronig-Penney
operator acting in $L^2(\RR)$ i.e. the point interaction Hamiltonian,
\begin{equation*}
  P_{u,\beta}:=-\dfrac{d^2}{dx^2}+\beta\sum_{k\in\ZZ}\delta(\cdot-k u).
\end{equation*}
The operator can be correctly defined through the associated sesquilinear form
\[
\langle f,  P_{u,\beta} g\rangle =\langle f', g'\rangle_{L^2}+
\beta \sum_{k\in\ZZ} \overline{f(ku)}\, g(ku), \quad f,g\in H^1(\RR),
\]
and is unitarily equivalent to the operator $H(l_u,\beta)$ for $d=1$,
where $l_u$ is the constant function, $l_u(e)\equiv u$ for all
$e\in\cE$.  It is well known \cite{AGHH} that
\begin{equation}
  \label{eq:1}
  \spec P_{u,\beta} =\big\{ k^2: \Im k\ge 0,\, \cos ku
  +\dfrac{\beta}{2k}\,\sin ku\in[-1,1] \big\}.
\end{equation}
In particular, each Dirichlet eigenvalue $(\pi n)^2/u^2$,
$n=1,2,\dots$, is a spectral edge, and the bands depend continuously
on both $\alpha$ and $u$.  As follows from the general theory of
random operators \cite{CL,PF}, one has
\begin{equation}
  \label{eq:2}
  \Sigma\equiv \Sigma(\alpha):=\bigcup_{u\in[l_\mi,l_\ma]} \spec H(l_u,\alpha).
\end{equation}
On the other hand, as shown in \cite{KP}, one has the identity $\spec
H(l_u,\alpha):=\spec P_{u,\alpha/d}$. Hence the almost sure spectrum $\Sigma(\alpha)$ of
$H^\omega(\alpha)$ is a union of bands, and the bottom of the spectrum
is given by
\begin{equation}
\label{eq-infspec}
  \inf \Sigma(\alpha) =\begin{cases}
    k^2:\, k \in (0,\pi/l_\ma) \text{ and } \, \cos kl_\ma
    +\dfrac{\alpha}{2kd}\,\sin kl_\ma=1, & \alpha>0,\\ 
    0, & \alpha=0,\\
    -k^2: k>0 \text{ and } \, \cosh kl_\mi +\dfrac{\alpha}{2kd}\,\sinh
    kl_\mi=1, & \alpha<0.
  \end{cases}
\end{equation}
Define the set
\begin{equation}
  \label{eq-delta}
  \Delta:=\bigcup_{n\in\ZZ}\Big[\dfrac{\pi^2 n^2}{l_\ma^2},\,
  \dfrac{\pi^2 n^2}{l_\mi^2}\Big].
\end{equation}
The set consists of the spectra of the operator $H(l_u)$ with
Dirichlet boundary condition (which formally corresponds to
$\alpha=\infty$) at each vertex when $u$ ranges over the support of
the random variables $(l^\omega_e)_{e\in\mathcal{E}}$ and the point $0$.
Our main result is
\begin{theorem}
  \label{th1}
  \textup{(a)} Let $\alpha\ne 0$ and $E_0$ be an edge of
  $\Sigma(\alpha)\cap (0,+\infty)$ that is not contained in $\Delta$.
  Then, there exists $\varepsilon>0$ such that the spectrum of
  $H^\omega$ in $(E_0-\varepsilon,E_0+\varepsilon)\cap\Sigma$ is
  almost surely dense pure point and the corresponding eigenfunctions
  are exponentially decaying.
  
  \noindent\textup{(b)} There exists $a>1$ and $\varepsilon>0$ such that 
  for $\alpha\in(-\infty, -a)\cup (-1/a,0)$ the spectrum of
  $H^\omega(\alpha)$ in $[\inf \Sigma,\inf \Sigma+\varepsilon$) is
  almost surely dense pure point with exponentially decaying
  eigenfunctions.
  
\end{theorem}
An immediate corollary of the above theorem and the equalities \eqref{eq:2}
is \begin{theorem}
  \label{th2} There exist $a>1$ and $\varepsilon>0$ such that
  for $\alpha\in (-\infty, -a)\cup (-1/a,0)\cup (0,+\infty)$ the
  spectrum of $H^\omega$ in $[\inf \Sigma,\inf \Sigma+\varepsilon)$ is
  almost surely dense pure point with exponentially decaying
  eigenfunctions.
\end{theorem}

\begin{rem} 
  An elementary analysis of the condition \eqref{eq:2} shows that, if
  $l_{\max}-l_{\min}$ is sufficiently small and $\alpha\ne0$,
  there are band edges of $\Sigma$ in $(0,+\infty)\setminus\Delta$.
   
  Our results do not establish the localization for the important case
  $\alpha=0$. In this case, both the Wegner estimate near
  $\inf\Sigma=0$ and the initial scale estimate (see
  theorems~\ref{th-wegner} and~\ref{th-ini} below) fail to hold. The
  operator $H(l^\omega,0)$ is analogous to the acoustic operator (see
  e.g.~\cite{MR2387562} and references therein); for this operator, at
  least in dimension one, it is known that localization does not hold
  in its strongest form at the bottom of the spectrum
  (see~\cite{MR714614}).
  In the case $\alpha=0$, the reduced operator
  we use to study the random quantum graph is a discrete version of
  the acoustic operator. 
  
  As in~\cite{FKP}, the assumptions of theorems~\ref{th1}
  and~\ref{th2} also imply dynamical localization (see remark 7 in
  ~\cite{FKP}).
\end{rem}

\section{Multiscale analysis and finite-volume operators}
\label{sec:mult-analys-finite}

Let $\Lambda$ be a subset of edges from $\cE_d$. Denote $
\cV_\Lambda:=\{\iota e:\, e\in\Lambda\}\cup \{\tau e:\, e\in\Lambda\}$
and consider the graph
$\Gamma^d_\Lambda:=\big(\cV_\Lambda,\Lambda\big)$.  Note that this
graph has no isolated vertices. We will call the operator
$H_\Lambda(l,\alpha):=H(\Gamma^d_\Lambda,l,\alpha)$ the finite-volume
Hamiltonian associated to $\Lambda$.  For random operators with random
length functions $l^\omega$, we write
$H^\omega_\Lambda(\alpha):=H_\Lambda(l^\omega,\alpha)$

In what follows, we consider Hamiltonians associated with finite cubes
$\Lambda=\Lambda(n)$ constructed as follows: take $n\in\NN$ and denote
by $\Lambda(n)$ the set of edges $e$ such that at least one of the
vertices $v\in \{\iota e,\tau e\}$ satisfies $|v|\le n$; for the
corresponding set of vertices, we write $\cV(n):=\cV_{\Lambda(n)}$.

As mentioned previously, the Anderson localization for random
operators can be established using a certain iterative procedure
called the multiscale analysis.  In order to start the multiscale
analysis one needs to verify the validity of several conditions for a
fixed interval $I\subset \RR$, which then imply the localization in
some subset of $I$ in various settings, see e.g.
\cite[Section~3.2]{St}.

The first group of conditions uses very few information on the nature
of random interactions, i.e. whether one has random edge length,
random potential on the edges or the random coupling constants etc.
and usually need only some uniform bounds for the random variables.
These conditions are as follows:
\begin{itemize}
  \setlength{\itemsep}{-2pt}
\item[(a)] the finite-volume Hamiltonians $H_\Lambda^\omega$ and
  $H_{\Lambda'}^\omega$ corresponding to any two non-overlapping
  finite sets of edges $\Lambda,\Lambda'\subset\cE_d$ are independent,
\item[(b)] the finite-volume operator obeys a Weyl estimate for the
  eigenvalues in $I$, i.e. there exists a constant $C>0$ such that the
  number of the eigenvalues of $H^\omega_n$ in $I$ can be estimated from
  above by $C n^d$ for all $n\in\ZZ$ and almost all $\omega\in\Omega$,
\item[(c)] there exists a geometric resolvent inequality which
  provides some uniform bounds for the resolvents of finite-volume
  operators in terms of the operators corresponding to smaller finite
  volumes,
\item[(d)] a generalized spectral theorem (Schnol-type theorem). This
  means that the existence of a non-trivial solution $f$ to $H^\omega
  f=Ef$ (i.e. $-f''_e=E f_e$ on each edge and the boundary conditions
  at the vertices are satisfied) with a suitable bound an infinity
  implies $E\in\spec H^\omega$ (for $E\in I$) and, moreover, the
  spectrum of $H^\omega$ in $I$ is the closure of the values $E\in I$
  with the above property.
\end{itemize}
Depending on the concrete problem, these conditions can weakened, see
e.g. \cite[Section~4]{Klein}. Note that the first condition (a) is
trivially satisfied in our case. The conditions (b) and (c) were shown
in \cite{EHS} for equilateral lattices, and the proof goes in our case
with minor modifications for any interval $I$. The generalized
spectral theorem, the condition (d), holds in any interval $I$ as well
due to the results of \cite{BLS} (see also \cite{LSS} for
generalizations).

The second group of conditions are very sensible to the way the
randomness enters the system. These two conditions are
\begin{itemize}
\item[(e)] The Wegner estimate showing that the probability for
  $H^\omega_\Lambda$ to have an eigenvalue in the
  $\varepsilon$-neighborhood of $E\in I$ can be globally bounded by $C
  \varepsilon^a |\Lambda|^{bd}$ with some $a,b\ge 1$.
\item[(f)] initial scale estimate showing that that the probability
  for $H^\omega_\Lambda$ to have an eigenvalue in the
  $|\Lambda|^{-a}$-neighborhood of \emph{some} $E_0\in I$ can be
  bounded by $|\Lambda|^{-b}$ with some suitable $a,b\ge 0$ (which
  depend on the dimension and other parameters).
\end{itemize}
When all the above conditions are satisfied, the multiscale analysis
shows Anderson localization in a certain interval around the energy
$E_0$ taken from the condition (f).

Hence, in the present Letter, we are interested in the Wegner estimate
and the initial scale estimates for our model, that is, (e) and (f).
They imply theorem \ref{th1} by the multiscale analysis.

\begin{theorem}[The Wegner estimate]
  \label{th-wegner}
  \textup{(a)} Let $I\subset(0,\infty)$ be an interval such that $\Bar
  I\cap \Delta=\emptyset$, then there exists a constant $C=C(I)>0$
  such that for any interval $J\subset I$, and any cube
  $\Lambda=\Lambda(n)$ there holds
  \begin{equation}
    \label{eq-th-weg}
    \PP\big\{\spec H^\omega_\Lambda(\alpha)\cap J
    \ne\emptyset\big\}\le C  |\Lambda|\,|J|.
  \end{equation}
  
  \noindent \textup{(b)} There exists $a>1$ such that for $\alpha\in
  (-\infty, -a)\cup (-1/a,0)$ the Wegner estimate also holds at
  negative energies near the bottom of the spectrum, i.e. there exists
  an interval $I$ with $I\ni \inf \Sigma(\alpha)$ and $C=C(I)>0$ such
  that for any interval $J\subset I$, and any cube
  $\Lambda=\Lambda(n)$ the estimate \eqref{eq-th-weg} holds.
  
\end{theorem}

\begin{theorem}[The initial scale estimate]
  \label{th-ini}
  Let $E$ be a spectral edge of $H^\omega(\alpha)$ which is not
  contained in the set $\Delta$ defined in \eqref{eq-delta}, then for
  each $\xi>0$ and $\beta\in(0,1)$ there exists $n^*=n^*(\xi,\beta)>0$
  such that, for $n\ge n^*$,
  \begin{equation*}
    \PP \{\dist(\spec H_{\Lambda(n)}^\omega(\alpha),E)\le
    n^{\beta-1}\}\le n^{-\xi}.
  \end{equation*}
\end{theorem}

\begin{rem}
  By \eqref{eq-infspec}, for $\alpha\ne 0$ the initial scale estimate
  is satisfied near the bottom of the spectrum $\inf \Sigma$, which is
  outside $\Delta$.

  When $\alpha=0$, independently of the random variables $l^\omega_e$
  and the set $\Lambda$, the constant function $f\equiv 1$ satisfies
  $H^\omega_\Lambda f=0$. Hence, both the Wegner estimate and the
  initial scale estimate fail for the energy $E=0$, and this is the
  only spectral edge; in this case, the almost sure spectrum is the
  positive half-line.\\
  Actually, in dimension $d=1$, the operator $H^\omega(0)$ is unitary
  equivalent to the free Laplacian and hence shows no Anderson localization
  (the spectrum is absolutely continuous). Hence, one has a certain
  similarity to the Schr{\"o}dinger operators with random vector
  potentials, where only localization near internal spectral edges is
  proved so far \cite{GHK}.
\end{rem}

We will prove both estimates, theorems \ref{th-wegner} and
\ref{th-ini} by exploiting a correspondence between the quantum graphs
and discrete operators.  A similar approach was used in \cite{FKP} for
quantum graphs with random coupling constants and more details on the
reduction can be found there.

Denote by $D^\omega_e$ the positive Laplacian with the Dirichlet
boundary conditions in $L^2[0,l^\omega_e]$ and set
$D^\omega_\Lambda:=\bigoplus_{e\in\Lambda} D^\omega_e$. Clearly,
\begin{equation*}
  \spec D^\omega_\Lambda=\bigcup_{e\in\Lambda} \spec D^\omega_e,\quad \spec
  D^\omega_e=\bigg\{ \Big(\dfrac{\pi k}{l^\omega_e}\Big)^2: \,
  k=1,2,\dots\bigg\}.
\end{equation*}
For $E\notin\spec D^\omega_\Lambda$ consider the operators
$M_\Lambda(l^\omega,E)$ acting on $\ell^2(\cV_\Lambda)$,
\begin{multline}
  \label{eq-mzgen}
  M_\Lambda(l^\omega, E) \varphi(v)= \sqrt{E} \, \Big(
  \sum_{e\in\Lambda: \iota e= v} \dfrac{1}{\sin l^\omega_e \sqrt E} \,
  \varphi(\tau e) + \sum_{e\in\Lambda: \tau e= v} \dfrac{1}{\sin
    l^\omega_e \sqrt E} \,
  \varphi(\iota e)\\
  {}-\sum_{e\in\Lambda:v\sim e} \cot l^\omega_e \sqrt E \,\varphi(v)
  \Big).
\end{multline}
Here and in what follows, the continuous branch of the square root is
fixed by the condition $\Im\sqrt E\ge 0$ for $E\in \RR$.

The meaning of the operators $M_\Lambda (l^\omega,E)$ is as follows.
Consider the equation $H^\omega_\Lambda f =E f$ for $E\notin \spec
D^\omega$.  On each edge $e\in\Lambda$, $f_e$ satisfies the Dirichlet
problem $-f_e''=E f_e$, $f_e(0)= f(\iota e)$, $f_e(l_e^\omega)=f(\tau
e)$ (see \eqref{eq-cont1} for the definition of the values $f(v)$).
Therefore,
\begin{equation*}
f_e(t)=f(\iota e) \dfrac{\sin \sqrt E(l^\omega -t)}{\sin \sqrt E l^\omega}
+ f(\tau e) \dfrac{\sin \sqrt E t}{\sin \sqrt E l^\omega}.  
\end{equation*}
Substituting this representation into the boundary conditions
\eqref{eq-cont2} yields $M_\Lambda (l^\omega,E) f_\Lambda=\alpha
f_\Lambda$ where $f_\Lambda=\big(f(v)\big)_{v\in {\cV_\Lambda}}$. For
the complete graph, $\Lambda=\cE^d$, we simply write $M(l^\omega, E)$
instead of $M_\Lambda(l^\omega, E)$.  The map $E\mapsto
M_\Lambda(l^\omega,E)$ is obviously analytic outside $\spec D^\omega$.
The following characterization of the spectrum of
$H^\omega_\Lambda(\alpha)$ shown in \cite{KP2} will be the key to our
analysis:
\begin{itemize}
\item an energy $E\notin\spec D^\omega_\Lambda$ is in the spectrum of
  $H^\omega_\Lambda(\alpha)$ if and only if $\alpha\in \spec
  M_\Lambda(l^\omega,E)$
\item for each such $E$, one has $\dim\ker \big(H^\omega_\Lambda
  (\alpha)-E\big)=\dim\ker \big(M_\Lambda(l^\omega,E)-\alpha\big)$.
\end{itemize}

For infinite $\Lambda$, one needs to use the self-adjoint extension
theory \cite{BGP}; for finite $\Lambda$ this relation has been known
for a long time, see e.g.  \cite{jvb,exdual}. 

We note that similar relations between quantum graphs and discrete
operators exist for more general boundary conditions at the vertices,
but the corresponding reduced operators $M(E)$ become much more
complicated, see \cite{OP}.

\section{Proof of theorem \ref{th-wegner} (Wegner estimate)}
\label{sec-weg}

As noted previously, one has
\begin{equation}
  \PP\big\{\spec H^\omega_\Lambda(\alpha)\cap J \ne\emptyset\big\}=
  \PP\big\{\exists E\in J:\, \alpha\in\spec M_\Lambda(l^\omega,E)\big\}.
\end{equation}
Note also that, for any $E_*\in I$, one can write
$M_\Lambda(l^\omega,E)=M_\Lambda(l^\omega,E_*)+(E-E_*)_\Lambda
B_\Lambda(l^\omega,E,E_*)$. Due to analyticity, one can find a
constant $b>0$ such that $\|B_\Lambda(l^\omega,E,E_*)\|\le b$ for all
$E_*,E\in I$ and $\Lambda\subset \cE$ and almost all
$\omega\in\Omega$.

On the other hand, the condition $\alpha\in\spec
M_\Lambda(l^\omega,E)$ implies the existence of a vector $\varphi\in
\ell^2(\cV_\Lambda)$, $\|\varphi\|=1$, such that
$\big(M_\Lambda(l^\omega,E)-\alpha\big)\varphi=0$.  Let $E_J$ be the
center of $J$. The estimate, for $E\in J$,
\begin{equation*}
  \big\|\big(M_\Lambda(l^\omega,E_J)-\alpha\big)\varphi\big\| \le
  \big\|\big(M_\Lambda(l^\omega,E)-\alpha\big)\varphi\big\| +
  |E-E_J|\cdot \big\| B_\Lambda(l^\omega,E,E_J)\varphi \big\|\le b |J|,
\end{equation*}
yields the inequality
\begin{equation}
  \label{eq-pej}
  \PP\big\{\spec H^\omega_\Lambda(\alpha)\cap J \ne\emptyset\big\}\le
  \PP\Big\{\dist \big(\spec M_\Lambda(l^\omega,E_J),\alpha\big)\le b|J| \Big\}.
\end{equation}
In what follows, we denote $E_J$ simply by $E$ to alleviate the
notation.

For $e\in\cE$, introduce the operators $P_1^e$, $P_2^e$, $I_e$ acting
on $\ell^2(\cV_\Lambda)$:
\begin{gather}
  P_1^e f(v)=\begin{cases}
    \dfrac{1}{2}\,\big(f(\iota e)+f(\tau e)\big), & v\in\{\iota e,\tau e\},\\
    0, & \text{otherwise},
  \end{cases}
  \\
  P_2^e f(v)=\begin{cases}
    \dfrac{1}{2}\,\big(f(\iota e)-f(\tau e)\big), & v=\iota e,\\[\bigskipamount]
    \dfrac{1}{2}\,\big(f(\tau e)-f(\iota e)\big), & v=\tau e,\\
    0, & \text{otherwise},
  \end{cases}
  \\
  I^e f(v)=\begin{cases}
    f(v), & v\in\{\iota e,\tau e\},\\
    0, & \text{otherwise}.
  \end{cases}
\end{gather}
In terms of these operators, one has
\begin{equation*}
  M_\Lambda(l^\omega,E) =\sum_{e\in\Lambda} \Big( \dfrac{\sqrt{E}}{\sin
    l^\omega_e\sqrt E}\,(P_1^e-P_2^e) -\sqrt E \cot l^\omega_e \sqrt E
  \,I^e \Big)
\end{equation*}
and
\begin{equation*}
  \dfrac{\partial M_\Lambda(l^\omega,E)}{\partial l^\omega_e}= -\dfrac{E
    \cos l^\omega_e\sqrt E}{\sin^2 l^\omega_e\sqrt E}\,(P_1^e-P_2^e)
  +\dfrac{E}{\sin^2 l^\omega_e\sqrt E}\, I^e.
\end{equation*}
Consider first the part (a) of the theorem, i.e. the case $\Bar
I\subset (0,+\infty)$. As $\|P_1^e-P_2^e\|=1$ and $P_j^e I^e=P_j^e$
for $j\in\{1,2\}$, one has
\begin{equation*}
  -\cos l^\omega_e\sqrt E \,(P_1^e-P_2^e) + I^e\ge \big(1-|\cos
  l^\omega_e\sqrt E|\big) I^e.
\end{equation*}
As $I$ does not meet $\Delta$, there exist constants $c_1,c_2>0$ such
that
\begin{equation*}
  1-|\cos l_e\sqrt E|\ge c_1 \text{ and } \dfrac{E}{\sin^2
    l^\omega_e\sqrt E}\ge c_2 \quad \text{for all } e\in \cE \text{ and
  } E\in I \text{ and a.e. } \omega\in\Omega.
\end{equation*}
Hence,
\begin{equation*}
  \dfrac{\partial M_\Lambda(l^\omega,E)}{\partial l_e^\omega} \geq c_1
  c_2 I^e \quad \text{for all } e\in \cE \text{ and } E\in I
\end{equation*}
so that
\begin{equation*}
  \sum_{e\in\cE} \dfrac{\partial M_\Lambda(l^\omega,E)}{\partial
    l_e^\omega} \geq c_1 c_2 \sum_{e\in\cE} I^e \geq \beta\, \text{id},
  \quad \beta=c_1c_2>0
\end{equation*}
or
\begin{equation*}
  D_\Lambda M_\Lambda(l^\omega,E)\ge \beta\, \text{id}\quad \text{
    with }\quad
  D_\Lambda:=\sum_{e\in\Lambda} \dfrac{\partial}{\partial l_e}.
\end{equation*}

Let $E^\omega_\Lambda(a,b)$ denote the spectral projection of
$M_\Lambda(l^\omega,E)$ onto the interval $(a,b)$. There holds
\begin{multline}
  \label{eq-mtr}
  \#\left[\spec M_\Lambda(l^\omega,E)\cap \big(\alpha-
    b|J|,\alpha+b|J|\big)\right]\\
  {}=\tr E_\Lambda^\omega(\alpha- b|J|,\alpha+b|J|\big) = \tr
  \left[\int_{\alpha- b|J|}^{\alpha+b|J|} \partial_\lambda
    \chi_{(-\infty,0]}(M_\Lambda(l^\omega,E_J)-\lambda)d\lambda\right].
\end{multline}
On the other hand, one has
\begin{multline*}
  -\tr\left[ D_\Lambda
    \chi_{(-\infty,0]}\big(M_\Lambda(l^\omega,E)-\lambda\big)\right]\\
  =\tr\left[\partial_\lambda
    \chi_{(-\infty,0]}(M_\Lambda(l^\omega,E)-\lambda) D_\Lambda
    M_\Lambda(l^\omega, E)\right] \ge \beta\tr\left[ \partial_\lambda
    \chi_{(-\infty,0]}\big(M_\Lambda(l^\omega,E)-\lambda\big)\right].
\end{multline*}
The last estimate is possible as both operators under the trace sign
are non-negative. Hence,
\begin{equation*}
  \tr\left[\partial_\lambda
    \chi_{(-\infty,0]}\big(M_\Lambda(l^\omega,E)-\lambda\big)\right] \le
  \beta^{-1}\, \tr\left[ \sum_{e\in \cE} -\partial_e
    \chi_{(-\infty,0]}\big(M_\Lambda(l^\omega,E)-\lambda\big)\right],
\end{equation*}
where we denoted for brevity $\partial_e:=\partial/\partial
l^\omega_e$, and
\begin{equation*}
  \tr\left[E^\omega_\Lambda\big(\alpha- b|J|,\alpha+b|J|\big)\right] \le
  \beta^{-1}\, \int_{\alpha- b|J|}^{\alpha+b|J|} \sum_{e\in \cE} \tr
  \left[-\partial_e \chi_{(-\infty,0]}
    \big(M_\Lambda(l^\omega,E)-\lambda\big)\right]d\lambda.
\end{equation*}
Taking the expectation, one obtains
\begin{equation}
  \label{eq-eeg}
  \EE \tr E_\Lambda(\alpha- b|J|,\alpha+b|J|\big)
  \le \beta^{-1}\,\sum_{e\in \cE} \int_{l_\mi}^{l_\ma} \prod_{e'\ne e}
  \rho(l^\omega_{e'}) dl^\omega_{e'} 
  \int_{\alpha- b|J|}^{\alpha+b|J|}  G_e(E,\lambda,\omega) d\lambda
\end{equation}
where$M_{\Lambda,e}(l^\omega,l,E)$ is the operator
$M_\Lambda(l^\omega,E)$ with $l^\omega_e$ replaced by $l$ and
\begin{equation*}
  G_e(E,\lambda,\omega)= -\int_{l_\mi}^{l_\ma} \rho(l)
  \partial_l \tr\left[ \chi_{(-\infty,0]}
    \big(M_{\Lambda,e}(l^\omega,l,E)-\lambda\big)\right] dl.
\end{equation*}
As the density $\rho$ is Lipschitz continuous by assumption, one can
integrate by parts and obtain
\begin{equation*}
  G_e(E,\lambda,\omega)= -\rho(l)
  F_e(l,E,\lambda,\omega)\big|_{l=l_\mi}^{l=l_\ma} +
  \int_{l_\mi}^{l_\ma} \rho'(l) F_e(l,E,\lambda,\omega) dl,
\end{equation*}
where
\begin{equation*}
  F_e(l,E,\lambda,\omega):= \tr
  \Big[\chi_{(-\infty,0]}\big(M_{\Lambda,e}(l^\omega,l,E)-\lambda\big)-
  \chi_{(-\infty,0]}\big(M_{\Lambda,e}(l^\omega,l_\mi,E)-\lambda\big)\Big].
\end{equation*}
As $\partial_e M_{\Lambda}(l^\omega,E)$ is a rank-two operator, the
functions $F_e(l,E,\lambda,\omega)$ are uniformly bounded by $2$.
Hence, the functions $G_e$ are bounded as well, say $|G_e|\le G$ for
some $G>0$. Plugging this estimate into \eqref{eq-eeg}, one obtains
\begin{equation*}
  \EE \tr E_\Lambda(\alpha- b|J|,\alpha+b|J|\big) \le
  G\beta^{-1}\,\sum_{e\in \cE} \int_{\alpha- b|J|}^{\alpha+b|J|}
  d\lambda = C |\Lambda| |J|, \quad C:= \dfrac{2bG}{\beta}>0.
\end{equation*}
It remains to observe that
\begin{equation}
  \label{eq-maj}
  \PP\Big\{\dist \big(\spec M_\Lambda(l^\omega,E),\alpha\big)\le b|J| \Big\}\le
  \EE \tr E_\Lambda(\alpha- b|J|,\alpha+b|J|\big).
\end{equation}

Now, let us now prove part (b) of theorem 4. Below we assume
$\alpha<0$.  Take first an arbitrary interval $I=(-E_+,-E_-)$ with
$E_+>E_->0$.  Note that for $E<0$, it is more convenient to rewrite
\begin{equation*}
  M_\Lambda(l^\omega,E) =\sum_{e\in\Lambda} \Big(
  \dfrac{\sqrt{-E}}{\sinh l^\omega_e\sqrt{-E}}\,(P_1^e-P_2^e) -\sqrt{
    -E} \coth l^\omega_e \sqrt{- E} \,I^e \Big).
\end{equation*}
Then,
\begin{equation*}
  \partial_e M_\Lambda(l^\omega,E) = \dfrac{|E|}{\sinh^2
    l^\omega_e\sqrt{-E}}\,\Big[ I_e -\cosh
  l^\omega_e\sqrt{-E}(P_1^e-P_2^e) \Big],
\end{equation*}
and one has
\begin{equation*}
  F_\Lambda M_\Lambda(l^\omega,E)=
  M_\Lambda(l^\omega,E)+K_\Lambda(l^\omega,E)
\end{equation*}
with
\begin{equation*}
  F_\Lambda:=- \dfrac{1}{\sqrt{-E}}\, \sum_{e\in\Lambda} \tanh
  l^\omega_e \sqrt{-E}\, \dfrac{\partial}{\partial l^\omega_e}, \quad
  K_\Lambda(l^\omega,E):=\dfrac{1}{\sqrt{-E}}\,\sum_{e\in\Lambda} \tanh
  l^\omega_e \sqrt{-E}\, I^e.
\end{equation*}
Denote $\gamma=\dfrac{\tanh l_\mi\sqrt{E_-}}{\sqrt{E_+}} $, then one
has $K_\Lambda(l^\omega,E)\ge \gamma\,\text{id}$ for all $\Lambda$,
$E\in I$, and a.e. $\omega$. As in the case $\alpha>0$, one computes
\begin{multline*}
  -\tr\left[F_\Lambda \left(\chi_{(-\infty,0]}
      \big(M_\Lambda(l^\omega,E)-\lambda\big)\right)\right]\\
  =\tr\left[\partial_\lambda \chi_{(-\infty,0]}
    (M_\Lambda(l^\omega,E)-\lambda) F_\Lambda M_\Lambda(l^\omega, E)\right]\\
  {}=\tr\left[\partial_\lambda
    \chi_{(-\infty,0]}(M_\Lambda(l^\omega,E)-\lambda)
    M_\Lambda(l^\omega, E)\right] +\tr \left[\partial_\lambda
    \chi_{(-\infty,0]}
    (M_\Lambda(l^\omega,E)-\lambda) K_\Lambda(l^\omega, E)\right]\\
  {}\ge \tr \left[ \lambda\partial_\lambda
    \chi_{(-\infty,0]}(M_\Lambda(l^\omega,E)-\lambda)\right]
  +\gamma \tr \left[\partial_\lambda
    \chi_{(-\infty,0]}(M_\Lambda(l^\omega,E)-\lambda) \right]\\
  =
  \tr \left[(\gamma+\lambda)\partial_\lambda
    \chi_{(-\infty,0]}(M_\Lambda(l^\omega,E)-\lambda) \right].
\end{multline*}
Assume that 
\begin{equation}
       \label{eq-gamma-beta}
\lambda+\gamma\ge \beta>0 \text{ for } \lambda\in (\alpha-b|I|,\alpha+b|I|)\equiv (\alpha-b|E_+-E_-|,\alpha+b|E_+-E_-|),
\end{equation}
then
\[
  \beta \tr \left[\partial_\lambda \chi_{(-\infty,0]}
    (M_\Lambda(l^\omega,E)-\lambda)\right]
  {}\le \tr \left[\sum_{e\in\Lambda} \dfrac{\tanh l^\omega_e
      \sqrt{-E}}{\sqrt{-E}}\, \dfrac{\partial}{\partial l^\omega_e}
    \chi_{(-\infty,0]}\big(M_\Lambda(l^\omega,E)-\lambda\big)\right].
\]
Using this inequality and \eqref{eq-mtr} one can estimate
\begin{multline*}
  \EE \tr \left[E_\Lambda(\alpha- b|J|,\alpha+b|J|\big) \right]
=\tr \left[\int_{\alpha- b|J|}^{\alpha+b|J|}\partial_\lambda
    \chi_{(-\infty,0]}\big(M_\Lambda(l^\omega,E)-\lambda\big)\,d\lambda
  \right]\\
  \le \beta^{-1}\int_{\alpha- b|J|}^{\alpha+b|J|} \sum_{e\in\Lambda}
  \dfrac{\tanh l^\omega_e \sqrt{-E}}{\sqrt{-E}}\,
  \dfrac{\partial}{\partial l^\omega_e}\tr
  \chi_{(-\infty,0]}\big(M_\Lambda(l^\omega,E)-\lambda\big)  \,d\lambda\\
\end{multline*}
This is then estimated exactly as in the proof of the part (a).

Hence we need to chose the interval $I$ in such a way that
\eqref{eq-gamma-beta} is satisfied. Let $E:=-\inf\spec
\Sigma(\alpha)$.  If one has
$\dfrac{\tanh l_\mi \sqrt E}{\sqrt E}+\alpha>0$,
then \eqref{eq-gamma-beta} holds for $E_-< E < E_+$ with $|E_+-E_-|$ sufficiently small.
On the other hand, by \eqref{eq-infspec}, 
\begin{equation}
       \label{eq-galph2}
\alpha=-2d\sqrt E \tanh \dfrac{l_\mi\sqrt E}{2}.
\end{equation}
Therefore, it is sufficient to find values of $\alpha$ for which
\begin{equation*}
f(E):=\dfrac{\tanh l_\mi \sqrt E}{\sqrt E}-2d\sqrt E \tanh \dfrac{l_\mi\sqrt E}{2}>0.  
\end{equation*}
A short computation shows that the sign of $f(E)$ coincides with the
sign of $\cosh^2 \dfrac{l_\mi \sqrt E}{2}- E d$ which is positive for
$E$ sufficiently close to $0$ as well as for $E$ sufficiently large,
and it remains to note that $E$ is a monotonous function of $\alpha$
due to \eqref{eq-galph2}. This completes the proof of theorem 4.

\section{Proof of theorem \ref{th-ini} (initial scale
  estimate)}\label{sec-lif}

As in the proof of theorem \ref{th-wegner}, one can show that, for
some $b>0$, there holds
\begin{equation}
  \label{eq-ini2}
  \PP \{\dist(\spec H_{\Lambda(n)}(l_\omega,\alpha),E)\le n^{\beta-1}\}\le
  \PP \{\dist(\spec M_{\Lambda(n)}(l_\omega,E),\alpha)\le  b n^{\beta-1}\}.
\end{equation}
Hence theorem \ref{th-ini} is a consequence of
\begin{equation}
  \label{eq-ini3}
  \PP \{\dist(\spec M_{\Lambda(n)}(l_\omega,E),\alpha)\le  b n^{\beta-1}\}
  \le n^{-\xi}.
\end{equation}
It is well known, see e.g. Section 2.2 in \cite{St} that, in order to
prove the estimates~\eqref{eq-ini3}, it is sufficient to show the
Lifshitz tail behavior for the integrated density of states for the
operator $M(l_\omega,E)$. Note that, if $E$ is an edge of the almost
sure spectrum of $H^\omega(\alpha)$, then $\alpha$ is an edge of the
almost sure spectrum of $M(l^\omega,E)$ (see e.g.\cite{FKP}). Hence,
it is sufficient to study the behavior of the integrated density of
states of $M(l^\omega,E)$ at the spectral edges.

The operator $M(l^\omega,E)$ is closely related to the random hopping
model considered in \cite{KN}; below, we use the very constructions
of~ \cite{KN} and~\cite{Na} to obtain the Lifshitz tails. The
integrated density of states in our case is defined by
\begin{equation*}
  k(t):=\lim_{n\to\infty} \dfrac{\#\{\lambda\in\spec
    M_{\Lambda(n)}(l_\omega,E):\,\lambda<t\}}{|\cV(n)|}.
\end{equation*}
This limit exists almost surely and $t\mapsto k(t)$ is non decreasing.
Let $[\mu_\mi,\mu_\ma]$ be the almost sure spectrum of
$M(l^\omega,E)$. Then, $k(t)=0$ for $t\leq \mu_{\min}$ and $k(t)=1$
for $t\geq \mu_{\max}$.
Denote also
\begin{equation*}
  b:=\sup_{l\in[l_\mi,l_\ma]} \Big|\dfrac{\sqrt{E}}{\sin l \sqrt{E}}\Big|.
\end{equation*}
By well-known arguments, see e.g. \cite[Section 2.2]{St}, in order to
prove \eqref{eq-ini3} it is sufficient to show
\begin{equation}
  \label{eq-dens}
  \lim_{\varepsilon\to 0+} \dfrac{\log
    \big|\log[1-k(\mu_\ma-\varepsilon)]
    \big|}{\log\varepsilon}\le -\dfrac{d}{2}
  \text{ and }
  \lim_{\varepsilon\to 0+} \dfrac{\log \big|\log
    k(\mu_\mi+\varepsilon)\big|}{\log\varepsilon}\le -\dfrac{d}{2}.
\end{equation}

For $n\in\NN$ define $M_n^\omega(E):=M(l^\omega_{n},E)$ where
$l^\omega_n(e)=l^\omega_n(e+\gamma)$ for $\gamma\in(2n+1)\ZZ^d$. By
the Floquet-Bloch theory, the operator $M_n^\omega(E)$ admits a
density of states, $k^\omega_n$, satisfying
\begin{equation*}
  k_n(E)=\dfrac{1}{(2\pi)^d}\int_{\big[-\frac{\pi}{2n+1},\frac{\pi}{2n+1}\big]}
  \#\spec M_n^\omega(E,\theta)\cap(-\infty,E)d\theta
\end{equation*}
where $M_n^\omega(E,\theta)$ differs from $M_n^\omega(E)$ only by an
operator of rank at most $C n^{d-1}$ with $C>0$ independent of $n$.
As suggested in \cite{KN}, in order to obtain \eqref{eq-dens}, it is
sufficient to show the analogous estimates with $k(E)$ replaced by
$\EE(k^\omega_n(E))$ uniformly in $n$ for sufficiently large $n$.
Then, as noted in \cite{Na} and applied in \cite{KN}, the latter
asymptotics can obtained directly from the following local energy
estimate which has been proved in \cite[Lemma~2.1]{KN}. Let
$a\in(0,b)$. For $\varphi\in l^2(\ZZ^d)$ one has
\begin{equation*}
  \langle \varphi, M_n^\omega \varphi \rangle \ge \langle \varphi,
  W^\omega_n \varphi\rangle +a\big\langle
  |\varphi|,H_0|\varphi|\big\rangle
\end{equation*}
where $H_0$ is the free Laplace operator in $(\ZZ^d)$,
\begin{equation*}
  H_0 u(n)=\sum_{n:|m-n|=1} \big( u(n)-u(m)\big),
\end{equation*}
and the potential $W_n^\omega$ is given by
\begin{equation*}
  W^\omega_n(v)=\sum_{e:v\sim v} \beta\Big(\dfrac{\sqrt{E}}{\sin
    l^\omega_e \sqrt E}\Big) +\sum_{e:v\sim e} \sqrt{E} \cot
  l^\omega_e\sqrt{E}
\end{equation*}
with
\begin{equation*}
  \beta(t):=\begin{cases}
    -|t|, & |t|\ge a,\\
    -a, & \text{otherwise.}
  \end{cases}
\end{equation*}
Then, as in~\cite{KN}, following the computations done in~\cite{Na},
one proves Lifshitz tails for $M(l^\omega,E)$ near $\mu_{\max}$ or
$\mu_{\min}$. This completes the proof of theorem~\ref{th-ini}.\qed

\section{Acknowledgments}
\label{sec:acknowledgments}

The authors thank Olaf Post and Ivan Veseli\'c for motivating
discussions.  The second named author was supported by the Marie Curie
Intra-European Fellowship (PIEF-GA-2008-219641).

\end{document}